\documentstyle[12pt,twoside]{article}
 
\def\d{\delta}
\def\e{\epsilon}                
\def\f{\phi}                    

\def\l{\lambda}
\def\m{\mu}
\def\n{\nu}

\def\r{\rho}                    
\def\s{\sigma}                  

\def\F{\Phi}

\def\O{\Omega}

\def\bo{{\raise.05ex\hbox{\large$\Box$}\:}}             
\def\cbo{{\,\raise-.15ex\Sc [\,}}                       
\def\pa{\partial}                                       
\def\de{\nabla}                                         
\def\su{\sum}                                           
\def\TH{{\raise.2ex\hbox{$\displaystyle \bigodot$}\mskip-4.7mu \llap H \;}}
\def\face{\hbox{\normalsize$\;\;\:{\raise.9ex\hbox{\oo n}\mskip-13mu \llap
        {${\buildrel{\hbox{\frtnrm ..}}\over\smile}$}}\:$}}     
\def\Face{{\raise.2ex\hbox{$\displaystyle \bigodot$}\mskip-2.2mu \llap {$\ddot
        \smile$}}}                                      
\def\Lhat{{\bf\rlap{\kern-.09em$\hat{\phantom L}$}L}}
\def\Lcheck{{\bf\rlap{\kern-.09em$\check{\phantom L}$}L}}
 
\def\sp#1{{}^{#1}}                              
\def\sb#1{{}_{#1}}                              
\def\svev#1{\left\langle #1\right\rangle}       

\def\leftrightarrowfill{$\mathsurround=0pt \mathord\leftarrow \mkern-6mu
        \cleaders\hbox{$\mkern-2mu \mathord- \mkern-2mu$}\hfill
        \mkern-6mu \mathord\rightarrow$}
\def\dvec#1{\vbox{\ialign{##\crcr
        \leftrightarrowfill\crcr\noalign{\kern-1pt\nointerlineskip}
        $\hfil\displaystyle{#1}\hfil$\crcr}}}           
\def\ddt#1{{\buildrel {\hbox{\LARGE .\kern-2pt.}} \over {#1}}}
 
\def\frac#1#2{{\textstyle{#1\over\vphantom2\smash{\raise.20ex
        \hbox{$\scriptstyle{#2}$}}}}}                   
\def\ha{\frac12}                                        
\def\sfrac#1#2{{\vphantom1\smash{\lower.5ex\hbox{\small$#1$}}\over
        \vphantom1\smash{\raise.4ex\hbox{\small$#2$}}}} 
\def\bfrac#1#2{{\vphantom1\smash{\lower.5ex\hbox{$#1$}}\over
        \vphantom1\smash{\raise.3ex\hbox{$#2$}}}}       
\def\afrac#1#2{{\vphantom1\smash{\lower.5ex\hbox{$#1$}}\over#2}}    
 
\def\boxes#1{
        \newcount\num
        \num=1
        \newdimen\downsy
        \downsy=-1.64ex
        \mskip-7.8mu
        \bo
        \loop
        \ifnum\num<#1
        \llap{\raise\num\downsy\hbox{$\bo$}}
        \advance\num by1
        \repeat}
\def\boxup#1#2{\newcount\numup
        \numup=#1
        \advance\numup by-1
        \newdimen\upsy
        \upsy=.82ex
        \mskip7.8mu
        \raise\numup\upsy\hbox{$#2$}}
 
\newskip\humongous \humongous=0pt plus 1000pt minus 1000pt
\def\caja{\mathsurround=0pt}

\newif\ifdtup
\def\panorama{\global\dtuptrue \openup2\jot \caja
        \everycr{\noalign{\ifdtup \global\dtupfalse
        \vskip-\lineskiplimit \vskip\normallineskiplimit
        \else \penalty\interdisplaylinepenalty \fi}}}
\def\li#1{\panorama \tabskip=\humongous                         
        \halign to\displaywidth{\hfil$\displaystyle{##}$
        \tabskip=0pt&$\displaystyle{{}##}$\hfil
        \tabskip=\humongous&\llap{$##$}\tabskip=0pt
        \crcr#1\crcr}}

\def\NP{Nucl. Phys. B\,}
\def\PL{Phys. Lett. B\,}
\def\PR{Phys. Rev. Lett.\,}

\def\CQG{Class. Quant. Grav.\,}

\def\ref#1{$\sp{#1]}$}

\parskip=\medskipamount                 
\def\baselinestretch{1.2}       
\thicklines                         
\def\title#1#2#3#4{
\begin{document}
        {\hbox to\hsize{#4 \hfill  #3}}\par
        \begin{center}\vskip.5in minus.1in {\Large\bf #1}\\[.5in minus.2in]{#2}
        \vskip1.4in minus1.2in {\bf ABSTRACT}\\[.1in]\end{center}
        \begin{quotation}\par}
\def\author#1#2{#1\\[.1in]{\it #2}\\[.1in]}

\def\AMIC{Aleksandar Mikovic\'c
\\[.1in]{\it Blackett Laboratory, Imperial College, Prince Consort Road, London
SW7 2BZ, UK}\\[.1in]}

\def\AMICIF{Aleksandar Mikovi\'c\,
\footnote{Work supported by MNTRS and Royal Society}
\\[.1in] {\it Blackett Laboratory, Imperial College, Prince Consort
Road, London SW7 2BZ, UK}\\[.1in]
and \\[.1 in]
{\it Institute of Physics, P.O. Box 57, 11001 Belgrade, Yugoslavia}
\footnote{Permanent address}\\ {\it E-mail:\, mikovic@castor.phy.bg.ac.yu}}

\def\AMSISSA{Aleksandar Mikovi\'c\,
\footnote{E-mail address: mikovic@castor.phy.bg.ac.yu}
\\[.1in] {\it SISSA-International School for Advanced Studies\\
Via Beirut 2-4, Trieste 34100, Italy}\\[.1in]
and \\[.1 in]
{\it Institute of Physics, P.O. Box 57, 11001 Belgrade, Yugoslavia}
\footnote{Permanent address}}

\def\AM{Aleksandar Mikovi\'c 
\footnote{E-mail address: mikovic@castor.phy.bg.ac.yu}
\\[.1in] {\it Institute of Physics, P.O.Box 57, Belgrade 11001, Yugoslavia}
\\[.1in]}

\def\AMsazda{Aleksandar Mikovi\'c 
\footnote{E-mail address: mikovic@castor.phy.bg.ac.yu}
and Branislav Sazdovi\'c \footnote{E-mail: sazdovic@castor.phy.bg.ac.yu}
\footnote{Work supported by MNTRS}
\\[.1in] {\it Institute of Physics, P.O.Box 57, Belgrade 11001, Yugoslavia}
\\[.1in]}

\def\AMVR{Aleksandar Mikovi\'c\,
\footnote{E-mail address: mikovic@castor.phy.bg.ac.yu}
\\[.1in] 
{\it Institute of Physics, P.O. Box 57, 11001 Belgrade, Yugoslavia}
\\[.2in]
Voja Radovanovi\'c \\[.1 in]
{\it Faculty of Physics, P.O. Box 550, 11001 Belgrade, Yugoslavia}}

\def\AMCVR{Aleksandar Mikovi\'c
\footnote{Permanent address: Institute of Physics, P.O. Box 57, 11001 
Belgrade, Yugoslavia}\footnote{E-mail: mikovic@fy.chalmers.se, 
mikovic@castor.phy.bg.ac.yu}
\\
{\it Institute of Theoretical Physics, Chalmers University of Technology,
S-412 96 Goteborg, Sweden}\\[.1in]
and
\\[.1in]
Voja Radovanovi\'c
\footnote{E-mail: rvoja@rudjer.ff.bg.ac.yu} \\
{\it Faculty of Physics, P.O. Box 550, 11001 Belgrade, Yugoslavia}}

\def\AMVVR{Aleksandar Mikovi\'c
\footnote{On leave from Institute of Physics, P.O. Box 57, 11001 
Belgrade, Yugoslavia}
\footnote{Supported by Comissi\'on Interministerial de Ciencia y Tecnologia}
\footnote{E-mail: mikovic@lie1.ific.uv.es}
\\
{\it Departamento de Fisica Te\'orica and IFIC, Centro Mixto Universidad
de Valencia-CSIC, Facultad de Fisica, Burjassot-46100, Valencia, Spain}
\\[.1in]
Voja Radovanovi\'c
\footnote{E-mail: rvoja@rudjer.ff.bg.ac.yu} \\
{\it Faculty of Physics, P.O. Box 368, 11001 Belgrade, Yugoslavia}}

\def\AMV{Aleksandar Mikovi\'c
\footnote{On leave from Institute of Physics, P.O. Box 57, 11001 
Belgrade, Yugoslavia}
\footnote{Supported by Comissi\'on Interministerial de Ciencia y Tecnologia}
\footnote{E-mail: mikovic@lie1.ific.uv.es}
\\
{\it Departamento de Fisica Te\'orica and IFIC, Centro Mixto Universidad
de Valencia-CSIC, Facultad de Fisica, Burjassot-46100, Valencia, Spain}}

\def\endtitle{\par\end{quotation}\vskip3.5in minus2.3in\newpage}
 
 
\def\endabstract{\par\end{quotation}
        \renewcommand{\baselinestretch}{1.2}\small\normalsize}
 
 
\def\xpar{\par}                                         

\def\letterhead{
        \centerline{\large\sf INSTITUTE OF PHYSICS}
        \centerline{\sf P.O.Box 57, 11001 Belgrade, Yugoslavia}
        \rightline{\scriptsize\sf Dr Aleksandar Mikovi\'c}
        \vskip-.07in
        \rightline{\scriptsize\sf Tel: 11 615 575}
        \vskip-.07in
        \rightline{\scriptsize\sf E-mail: MIKOVIC@CASTOR.PHY.BG.AC.YU}}

\def\sig#1{{\leftskip=3.75in\parindent=0in\goodbreak\bigskip{Sincerely yours,}
\nobreak\vskip .7in{#1}\par}}

\def\ssig#1{{\leftskip=3.75in\parindent=0in\goodbreak\bigskip{}
\nobreak\vskip .7in{#1}\par}}

 
\def\ree#1#2#3{
        \hfuzz=35pt\hsize=5.5in\textwidth=5.5in
        \begin{document}
        \ttraggedright
        \par
        \noindent Referee report on Manuscript \##1\\
        Title: #2\\
        Authors: #3}
 
 
\def\start#1{\pagestyle{myheadings}\begin{document}\thispagestyle{myheadings}
        \setcounter{page}{#1}}
 
 
\catcode`@=11
 
\def\ps@myheadings{\def\@oddhead{\hbox{}\footnotesize\bf\rightmark \hfil
        \thepage}\def\@oddfoot{}\def\@evenhead{\footnotesize\bf
        \thepage\hfil\leftmark\hbox{}}\def\@evenfoot{}
        \def\sectionmark##1{}\def\subsectionmark##1{}
        \topmargin=-.35in\headheight=.17in\headsep=.35in}
\def\ps@acidheadings{\def\@oddhead{\hbox{}\rightmark\hbox{}}
        \def\@oddfoot{\rm\hfil\thepage\hfil}
        \def\@evenhead{\hbox{}\leftmark\hbox{}}\let\@evenfoot\@oddfoot
        \def\sectionmark##1{}\def\subsectionmark##1{}
        \topmargin=-.35in\headheight=.17in\headsep=.35in}
 
\catcode`@=12
 
\def\sect#1{\bigskip\medskip\goodbreak\noindent{\large\bf{#1}}\par\nobreak
        \medskip\markright{#1}}
\def\chsc#1#2{\phantom m\vskip.5in\noindent{\LARGE\bf{#1}}\par\vskip.75in
        \noindent{\large\bf{#2}}\par\medskip\markboth{#1}{#2}}
\def\Chsc#1#2#3#4{\phantom m\vskip.5in\noindent\halign{\LARGE\bf##&
        \LARGE\bf##\hfil\cr{#1}&{#2}\cr\noalign{\vskip8pt}&{#3}\cr}\par\vskip
        .75in\noindent{\large\bf{#4}}\par\medskip\markboth{{#1}{#2}{#3}}{#4}}
\def\chap#1{\phantom m\vskip.5in\noindent{\LARGE\bf{#1}}\par\vskip.75in
        \markboth{#1}{#1}}
\def\refs{\bigskip\medskip\goodbreak\noindent{\large\bf{REFERENCES}}\par
        \nobreak\bigskip\markboth{REFERENCES}{REFERENCES}
        \frenchspacing \parskip=0pt \renewcommand{\baselinestretch}{1}\small}
\def\unrefs{\normalsize \nonfrenchspacing \parskip=medskipamount}
\def\Item{\par\hang\textindent}
\def\Itemitem{\par\indent \hangindent2\parindent \textindent}
\def\makelabel#1{\hfil #1}
\def\topic{\par\noindent \hangafter1 \hangindent20pt}
\def\Topic{\par\noindent \hangafter1 \hangindent60pt}

\include{psfig}
\title{General Solution for Self-Gravitating Spherical Null Dust}
{\AMV}{FTUV/97-24, IFIC/97-24}{May 1997}

We find the general solution of equations of motion for self-gravitating 
spherical null dust as a perturbative series in powers of the outgoing matter 
energy-momentum tensor,
with the lowest order term being the Vaidya solution for the ingoing matter. 
This is done by 
representing the null-dust model as a 2d dilaton gravity theory, and by using a 
symmetry of a pure 2d dilaton gravity to fix the gauge. Quantization of this
solution would provide an effective metric which includes the back-reaction
for a more realistic black hole evaporation model than the 
evaporation models studied previously.


\endtitle

Two-dimensional (2d) dilaton gravity theories have turned out to be very useful
toy models of black hole formation and evaporation \cite{rev}. They are also 
relevant for four-dimensional (4d) black holes, since a spherically symmetric 
scalar field collapse can be described by a 2d dilaton gravity model
$$S_0 ={1\over2}\int d^2x\sqrt{-g}\left[e^{-2\F}\left(R+2 (\nabla_\m
\F )^2 + 2e^{2\phi}\right)-{G\over2}e\sp{-2\F}
(\nabla_\m f )^2\right]\quad,\eqno(1)$$
where $G$ is the Newton constant and the 4d line element $ds_4$ is related to 
the 2d line element $ds$ by
$$ds\sb 4\sp 2 =ds^2 +  e^{-2\F} d\Omega^2 \quad.\eqno(2)$$
$R$ is a 2d scalar curvature associated with a 2d metric $g_{\mu\nu}$,
$\nabla_\m$ is the corresponding covariant derivative,
$\F$ is the dilaton field, $f$ is the scalar field and
$d\O$ is a two-sphere line element. Quantization of (1)
would provide us with a semi-classical metric which would include the 
back-reaction.
However, the progress is hindered by the absence of explicit solutions of
the classical equations of motion. 
In the case of 2d black holes described by the CGHS model \cite{cghs}, 
the analog of (1) is 
exactly solvable, and by quantizing the solution, one can obtain an
effective semi-classical metric up to any finite order in matter loops
\cite{m95,m96,mr}. Since the matter in the CGHS model is 2d conformally coupled,
this motivates us to consider a modification of (1) 
$$S={1\over2}\int d^2x\sqrt{-g}\left[e^{-2\F}\left(R+2 (\nabla_\m
\F )^2 + 2e^{2\phi}\right)-{G\over2}
(\nabla_\m f )^2\right]\quad,\eqno(3)$$
so that $f$ will obey a free-field equation of motion 
in the conformal
gauge, and consequently the general solution could be found more easily.
Action (3) describes the dynamics of
spherically symmetric self-gravitating null-dust (SSND). For 
purely ingoing matter, the solution of the equations of motion is given by
the Vaidya metric \cite{Vaidya}
$$ds^2=-\left(1-{2m(v)\over r}\right)dv^2 + 2dvdr \quad,\eqno(4)$$
where $r =\exp(-\F)$, $G=1$ and
$$ {dm (v)\over dv} = T_{vv} (v) = {1\over2}(df/ dv)\sp 2 
\quad.\eqno(5)$$
An analytic model of black hole evaporation based on the 
quantization of the Vaidya solution has
been studied in \cite{cmns}, and a qualitative agreement with
the numerical results of \cite{Parentani} has been found.
However, in order to fully take into account the back-reaction in the
analytic approach, one needs the most
general solution for which the outgoing matter is also present 
\cite{hiscock,cmns}.

In order to find the general solution of the SSND model, we rewrite the 
action (3) as
$$S={1\over2}\int d^2x\sqrt{-{\tilde g}}\left[ {\tilde R}\f 
 + V(\f) - \ha ({\tilde\nabla}_\m f )^2\right]\quad,\eqno(6)$$
where $\f =e\sp{-2\F} =r\sp 2$, $\tilde g_{\m\n} = r g_{\m\n}$ and $V =2/r$. 
We do this
in order to establish the connection with a generic 2d dilaton gravity model,
which can be represented by the action of the form (6). For example, the
CGHS model is given by $V = 4\l\sp 2$, where $\l$ is a 2d cosmological constant. 
The equations of motion are given by
$$ \de_\m \de_\n \f -\ha g_{\m\n} V = -T_{\m\n}\quad, \eqno(7)$$
$$ R + {dV\over d\f} = 0\quad, \quad\Box f = 0 \quad,\eqno(8)$$
where we have omitted the tildas, $\Box = g\sp{\m\n}\de_\m \de_\n$ and
$$ T_{\m\n} = \pa_\m f \pa_\n f - \ha g_{\m\n} (\nabla_\m f)\sp 2 \quad.$$
In the conformal gauge $d{\tilde s}\sp 2 = -e\sp{2\r}dx\sp + dx\sp - $ one 
obtains
$$\pa_+ \pa_- \f = -\frac14 V e\sp{2\r} \quad,\eqno(9)$$
$$\pa_+\sp 2 \f -2\pa_+ \r \pa_+ \f = -(\pa_+ f )\sp 2 \quad,\eqno(10)$$
$$\pa_-\sp 2 \f -2\pa_- \r \pa_- \f = -(\pa_- f )\sp 2 \quad,\eqno(11)$$
$$\pa_+ \pa_- \r = -{1\over 8} {dV\over d\f} e\sp{2\r}\quad, \eqno(12)$$
and $\pa_+\pa_- f = 0$.
We want to solve the system (9-12) in analogy to the CGHS case, where
a gauge $\r =0$ can be chosen. This is possible since $\pa_+ \pa_- \r =0$
is a consequence of the equations of motion. 
In the SSND case $\r$ is not a free field, so that
we have to find an appropriate modification. We will use as the starting point
the free-field currents for the $f=0$ case \cite{cnt}
$$ j_1 = \int {d\f\over 2E + J(\f)}\quad, \eqno(13)$$
$$ \tilde j_2 = \log (\de\f)\sp 2 \quad,\eqno(14)$$
where $2E = (\de\f)\sp 2 - J(\f)$, $dJ/d\f = V$ and $E$ is a constant of motion.
$j$'s satisfy
$$ \Box j_1 = 0 \quad,\quad \Box \tilde j_2 + R =0 \quad. \eqno(15)$$
By going into the conformal gauge one
can see that the equations (15) imply that $j_1$ and $j_2 = \tilde j_2 -2\r$ 
are free fields. In the SSND case, when $f=0$ one has 
$ 2j_1 = r + 2M\log |r/2M - 1| $ and $ j_2 =\log |r - 2M | -2\r$,
where $M = -E/4$ is the black hole mass. The Schwarzschild solution is 
obtained for
$$2j_1 = \ha (v - u ) \quad,\eqno(16)$$
$$ j_2 = 0 \quad.\eqno(17)$$
The gauge choice (16) gives the familiar relation between $r$ and $u,v$
coordinates, while the gauge choice (17) is equivalent to
$$ds\sp 2 = -{e\sp{2\r}\over r} dudv = -(1-2M/r) dudv .\eqno(18)$$
Note that the gauge choice $j_2 =0$ gives a relation between $\r$ and $\f$,
which is of the type we are looking for, and hence we will concentrate on
it. 

When the matter is added, the relations (15) change as
$$\li{ \Box j_1 =& {\pa\sp 2 j_1 \over \pa E\sp 2} T\sp{\m\s}T\sb\m\sp\n
\de\sb\s \f \de\sb\n \f + \left( {\pa j_1 \over \pa E } -{4\over (\de \f)\sp 4}
\right) T\sb{\m\n} \de\sp\m \f \de\sp\n \f  \quad,\cr 
 \Box \tilde j_2 + R =& - {4\over (\de \f)\sp 4}
[T\sp{\m\s}T\sb\m\sp\n \de\sp\s \f \de\sp\n \f + 
V(\f)T\sb{\m\n} \de\sp\m \f \de\sp\n \f ] \cr 
 &+ {2\over (\de \f)\sp 2}T\sb{\m\n} \de\sp\m \f \de\sp\n \f \quad, &(19)\cr}$$
so that $j$'s are not free fields any more. The form of (19) and our remark 
about $j_2$ suggest that we look for a free field of the form
$$ j = j_2 + X \quad,\eqno(20)$$ 
where $X$ is to be determined from (19). This gives
$$\pa_+ \pa_- X = -  \left[ {T_{++}\over (\pa_+\f)\sp 2} +
{T_{--}\over (\pa_-\f)\sp 2} \right] \pa_+ \pa_- \f \quad. \eqno(21)$$
Equation (21) is valid for any potential $V$, and can be solved as
$$ X = X_0 + \log \left|{\pa_-\f\over\pa_+\f}\right| + 2 \int dx\sp - 
{T_{--}\over\pa_-\f} \quad,\eqno(22)$$
where $X_0$ is a free-field solution. 

In the SSND case $\f = r\sp 2$, $x\sp + = v$, $x\sp - = u$, 
and we choose $X_0 =0$, which fixes the gauge. 
It is convenient to use a generalized form of the Vaidya metric \cite{bard}
$$ ds\sp 2 = -e\sp{2\psi} F dv\sp 2 + 2 e\sp\psi dr dv \quad,\eqno(23)$$
which can be related to the conformal form of the metric via 
$ds\sp 2 = - C(u,v) du dv$ and $C = e\sp{2\r}/r$. 
Then the gauge choice $j =0$ is equivalent to
$$ \psi = 2\int du {T_{uu} \over \pa_u r\sp 2} \quad, \eqno(24)$$
where $T_{uu} =\ha T_{--}$. Note that we have not specified the limits of
$u$ integration in (24), which means that a constant of integration will
occur. Only when this constant is specified, the gauge will be completely 
fixed.

The equations which determine $r$ are (9) and (10),
while (11) and (12) are the consistency conditions for the gauge choice, 
which are satisfied by construction. Equation (9) becomes
$$\pa_u \pa_v r\sp 2 = e\sp\psi \pa_u r \quad,\eqno(25)$$
while (10) gives
$$\pa_v\sp 2 r\sp 2 -2\pa_v \r \pa_v r\sp 2 = -(\pa_v f )\sp 2 \quad,
\eqno(26)$$
where $\r$ is determined from the gauge choice as 
$$ 2\r = \log |-\pa_u r\sp 2 | + \psi \quad. \eqno(27)$$
Equation (11) follows from (27) and (24), while the equation (12) follows from
(27),(25) and (24). 
By integrating (25) with respect to $u$, we obtain
$$ \pa_v r = {1\over 2} \left( 1 - {2 m(v)\over r}\right) + {1\over 2r}
\int du (e\sp\psi - 1) \pa_u r \quad.\eqno(28)$$
$m (v)$ is a constant of integration, and can be determined from (26).
By inserting (28) into (26), and by using (27), we get
$$ T_{vv} = {dm\over dv} -{1\over 2} (1 - e\sp\psi -2r\pa_v\psi)\pa_v r
-{1\over 2} \pa_v \int du (e\sp\psi - 1) \pa_u r \quad.\eqno(29)$$

When $\psi =0$, (28) yields $F = 1 -2m/r$ and (29) gives the relation (5),
so that one recovers the Vaidya solution. When $\psi \ne 0$, the equations look
difficult. However, their form is such that a perturbative solution in 
$T_{uu}$ can
be easily found. Let us introduce an expansion parameter $\e$ by replacing
$T_{uu}$ with $\e T_{uu}$. Then we will seek a solution in the form
$$ r = r_0 + \e r_1 + \e\sp 2 r_2 + \cdots \quad.\eqno(30)$$
The expansion (30) then implies
$$ \psi = \e \psi_1 + \e\sp 2 \psi_2 + \cdots \quad,\eqno(31)$$
where
$$\psi_1 = \int du {T_{uu}\over r_0\pa_u r_0} \quad,\quad
\psi_2 = -\int du T_{uu}{\pa_u(r_1 r_0)\over (r_0 \pa_u r_0 )\sp 2} \quad, 
\quad...\quad. \eqno(32)$$
In general one should also take
$$ m(v) = m_0 (v) + \e m_1 (v) + \e\sp 2 m_2 (v) + \cdots \quad,\eqno(33)$$
although it is possible that the series in (33) gets truncated, 
like in the example we consider in this paper, where $m = \e m_1$. 
By inserting the expansions (30), (31) and (33) 
into the equation for $r$ (28), we get an infinite 
hierarchy of equations
$$\pa_v r_0 = {1\over 2} \left( 1 - {2 m_0\over r_0}\right)\quad,$$
$$ \pa_v r_1 =  { m_0\over r_0\sp 2} r_1 -{ m_1\over r_0}
+ {1\over 2r_0}\int du \psi_1 \pa_u r_0 \quad,$$
$$ \pa_v r_2 =  { m_0\over r_0\sp 2} r_2 -{ m_2\over r_0}
 -{ r_1\over r_0} \pa_v r_1
 + {1\over 2r_0}\int du \left[\left(\ha\psi_1\sp 2 + \psi_2 \right) \pa_u r_0 
+ \psi_1 \pa_u r_1\right], \quad...\,\, .\eqno(34)$$ 
The system (34) can be solved, since for every $n$ the equation for $r_n$ 
does not involve $r_k$ with $k > n$ and
each equation is a first order
linear differential equation for 
$r_n$, except for $n=0$, which is a non-linear first order differential
equation. Therefore by starting from $n=1$, one can write
an explicit solution for any $r_n$ in terms of $r_0$ and $T_{uu}$.
At each step several integration constant $C (u)$ and $C(v)$
will arise (due to $u$ and $v$ integrations), and these
constants can be determined from the constraint equation (29), 
exactly as in the
$T_{uu}=0$ case. This can be done because the equation (29) also decomposes 
into an infinite hierarchy of equations under the expansions (30), (31) and (33)
$$\li{ T_{vv} =& dm_0 /dv \quad,\cr
 0 =& dm_1/dv +\ha(\psi_1 + 2r_0 \pa_v \psi_1 )\pa_v r_0 - 
\ha\pa_v \int du\psi_1 \pa_u r_0 
\quad,\cr
0=& dm_2/dv +\ha(\psi_2 + \ha\psi_1\sp 2 + 2r_0 \pa_v \psi_2 + 
2r_1 \pa_v \psi_1 )\pa_v r_0 + \ha(\psi_1 + 2r_0 \pa_v \psi_1) \pa_v r_1\cr
&- \ha\pa_v \int du [(\psi_2 +\ha\psi_1\sp 2 ) \pa_u r_0 + \psi_1\pa_u r_1 ] 
\quad,\quad ...\quad.&(35)\cr}$$
The equations (35) will also determine the $m_n (v)$, provided we fix the
integration constants $\psi_n (v)$ which appear in (32). These are related to
the complete specification of the gauge, or equivalently, to the choice
of the $v$ coordinate, since a coordinate change $v=v(\tilde v )$ in (23)
gives
$$ \tilde\psi = \psi + \log {dv\over d\tilde v} \quad.\eqno(36)$$
At the end, one sets $\e =1$ and writes the solution as
$$r = r_0 +  r_1 +  r_2 + \cdots \quad,\quad \psi = \psi_1 + \psi_2 + \cdots 
\quad.\eqno(37)$$

These general features can be nicely illustrated on the example of a white
hole emiting a shock-wave. This is simply a shock-wave Vaidya solution where 
the $u$ and $v$ coordinates are interchanged. The corresponding spacetime is
described by the Penrose diagram of Fig.1. 
\begin{figure}[h]
\centerline{\psfig{figure=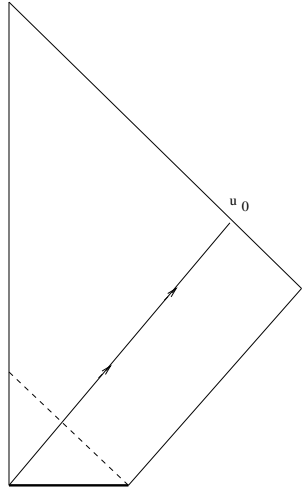}}
\caption{Penrose diagram of a white hole emiting a shock-wave.}
\end{figure}
In this case $T_{uu} \ne 0$, and
one can find an exact solution for $r=r(u,v)$, so that the expansions (30) and
(31) and the corresponding equations (34) and (35) can be checked. By taking
$T_{uu} = M \d (u - u_0 )$, we get
$$ ds\sp 2 = \left\{ \begin{array}{c}-du\sp 2 - 2du dr \quad u > u_0 \\ 
                    -(1 -2M/r)du\sp 2 - 2du dr \quad u < u_0\end{array}\right.
                     \quad.\eqno(38)$$
This can be rewritten as                    
$$ ds\sp 2 = \left\{ \begin{array}{c}-dv\sp 2 + 2dv dr \quad u > u_0 \\ 
                    -F e\sp{2\psi}dv\sp 2 + 2e\sp\psi dv dr \quad u < u_0
                    \end{array}\right.
                     \quad,\eqno(39)$$
where $F = 1 - 2M/r$,
$$ \psi = - \log \left| 1 - {4M\over v - u_0}\right| \eqno(40)$$
and $r =\ha (v - u)$ for $u>u_0$, while for $u<u_0$
$$ r + r_s \log |r/r_s - 1| = \ha (v-u) + r_s \log |(v-u_0 )/2r_s - 1| 
\quad,\eqno(41)$$
where $r_s = 2M$. The form of the solution, given by (40) and (41), is such that
$$ r = \su_{n\ge 0} r_s\sp n \tilde r_n \quad,\quad 
\psi = \su_{n\ge 1} r_s\sp n \tilde \psi_n \quad,\eqno(42)$$ 
which is of the form (37). More exactly, 
one can show
that $r_n = r_s\sp n \tilde r_n $ and $\psi_n = r_s\sp n \tilde \psi_n $ 
satisfy the equations (34) and (35) with $m = m_1 = M$. 
Also, starting from 
$n =2$, non-trivial integration constants $\psi_n (v)$ appear. 

In conclusion we can say that we have found a useful form of the general
solution, given by the expansions (37) and the equations (34) and (35).
The expansions (37) are clearly in powers of the outgoing energy-momentum 
tensor, and by truncating them at finite $n$ we obtain an explicit 
perturbative solution.
This form of the solution can be used to construct an 
effective quantum metric in the approximation of
a finite number of matter loops via the method of quantization of the classical 
solution 
\cite{m95,m96,mr,cmns}. The corresponding construction is going to be more 
involved than in the 2d case \cite{m95,m96,mr} or the Vaidya case \cite{cmns},
since the relation between the metric and the dilaton (radius)
is more complicated. For example, for the one-loop approximation one would take
$r = r_0 + r_1$ and $\psi = \psi_1$, with $T_{\m\n}$ replaced by 
$\svev{T_{\m\n}}$ evaluated in an appropriate quantum state. Note that 
in the one-loop case
one does not have to truncate the expansions (37) at $n=1$. By
including the higher-order terms, one extends the validity of the one-loop
approximation to a smaller radius.

Our solution can also serve as a good starting point for
finding approximate analytic solutions for the more realistic 
collapse described by the action (1).  

\sect{Acknowledgements}

\noindent I would like to thank Jose Navarro-Salas and Javier Cruz for useful 
discussions.

\end{document}